%% file: main.tex
\documentclass[sigconf]{acmart}
\usepackage{listings}
\usepackage{bm}
\usepackage{lipsum}
\usepackage{algorithm}
\usepackage{algpseudocode}
\usepackage{xcolor}
\usepackage{multicol}
\usepackage{multirow}
\usepackage{newtxmath}
\usepackage{caption}
\usepackage{graphicx}
\usepackage{subfigure}
\usepackage{amsmath}
\usepackage{balance}
\usepackage{marvosym}
\usepackage{enumitem}
\usepackage{svg}

\newcommand{\simtier}{\sc{SimTier}}
\newcommand{\make}{\sc{MAKE}}
\AtBeginDocument{%
  \providecommand\BibTeX{{%
    \normalfont B\kern-0.5em{\scshape i\kern-0.25em b}\kern-0.8em\TeX}}}

\copyrightyear{2024}
\acmYear{2024}
\setcopyright{acmlicensed}\acmConference[CIKM '24]{Proceedings of the 33rd ACM International Conference on Information and Knowledge Management}{October 21--25, 2024}{Boise, ID, USA}
\acmBooktitle{Proceedings of the 33rd ACM International Conference on Information and Knowledge Management (CIKM '24), October 21--25, 2024, Boise, ID, USA}
\acmDOI{10.1145/3627673.3680068}
\acmISBN{979-8-4007-0436-9/24/10}

\begin{document}

\title{Enhancing Taobao Display Advertising with Multimodal Representations: Challenges, Approaches and Insights}

\author{Xiang-Rong Sheng}
\authornote{Equal contribution.} 
\email{xiangrong.sxr@alibaba-inc.com}
\author{Feifan Yang} \authornotemark[1]
\email{yangfeifan.yff@alibaba-inc.com}
\author{Litong Gong} \authornotemark[1]
\email{gonglitong.glt@alibaba-inc.com}
\author{Biao Wang} \authornotemark[1]
\email{eric.wb@alibaba-inc.com}
\affiliation{%
  \institution{Alibaba Group}
  \city{Beijing}
  \country{China}
}

\author{Zhangming Chan}
\email{zhangming.czm@alibaba-inc.com}
\author{Yujing Zhang}
\email{jinghan.zyj@alibaba-inc.com}
\author{Yueyao Cheng}
\email{yueyao.syy@alibaba-inc.com}
\author{Yong-Nan Zhu}
\email{yongnan.zy@alibaba-inc.com}
\affiliation{%
  \institution{Alibaba Group}
  \city{Beijing}
  \country{China}
}

\author{Tiezheng Ge}
\email{tiezheng.gtz@alibaba-inc.com}
\author{Han Zhu}
\email{zhuhan.zh@alibaba-inc.com}\authornote{Corresponding author.} 
\author{Yuning Jiang}
\email{mengzhu.jyn@alibaba-inc.com}
\author{Jian Xu}
\email{xiyu.xj@alibaba-inc.com}
\author{Bo Zheng}
\email{bozheng@alibaba-inc.com}
\affiliation{%
  \institution{Alibaba Group}
  \city{Beijing}
  \country{China}
}

\renewcommand{\authors}{Xiang-Rong Sheng, Feifan Yang, Litong Gong, Biao Wang, Zhangming Chan, Yujing Zhang, Yueyao Cheng, Yong-Nan Zhu, Tiezheng Ge, Han Zhu, Yuning Jiang, Jian Xu, Bo Zheng}
\renewcommand{\shortauthors}{Xiang-Rong Sheng et al.}


\begin{abstract}
Despite the recognized potential of multimodal data to improve model accuracy, many large-scale industrial recommendation systems, including Taobao display advertising system, predominantly depend on sparse ID features in their models. 
In this work, we explore approaches to leverage multimodal data to enhance the recommendation accuracy. We start from identifying the key challenges in adopting multimodal data in a manner that is both effective and cost-efficient for industrial systems. 
To address these challenges, we introduce a two-phase framework, including: 1) the pre-training of multimodal representations to capture semantic similarity, and 2) the integration of these representations with existing ID-based models.
Furthermore, we detail the architecture of our production system, which is designed to facilitate the deployment of multimodal representations.
Since the integration of multimodal representations in mid-2023, we have observed significant performance improvements in Taobao display advertising system. We believe that the insights we have gathered will serve as a valuable resource for practitioners seeking to leverage multimodal data in their systems. 
\end{abstract}


\maketitle

\section{Introduction}
Traditionally, the recommendation models employed in Taobao's display advertising system, as with many other industrial systems, have largely relied on discrete IDs as features. Despite their widespread use, ID-based models have intrinsic drawbacks, such as the inability to capture the semantic information contained within multimodal data.

To address these issues, certain industrial systems have attempted to incorporate multimodal data into the ID-based models~\cite{Yang2023Courier,Singh2023SemanticID}. 
Typically, these approaches employ a two-phase framework, including 1) acquiring multimodal representations through either generic or scenario-specific pre-training, and 2) integrating these representations into the recommendation models. 

\begin{figure*}[!t]
\centering 
\includegraphics[width=0.9\textwidth]{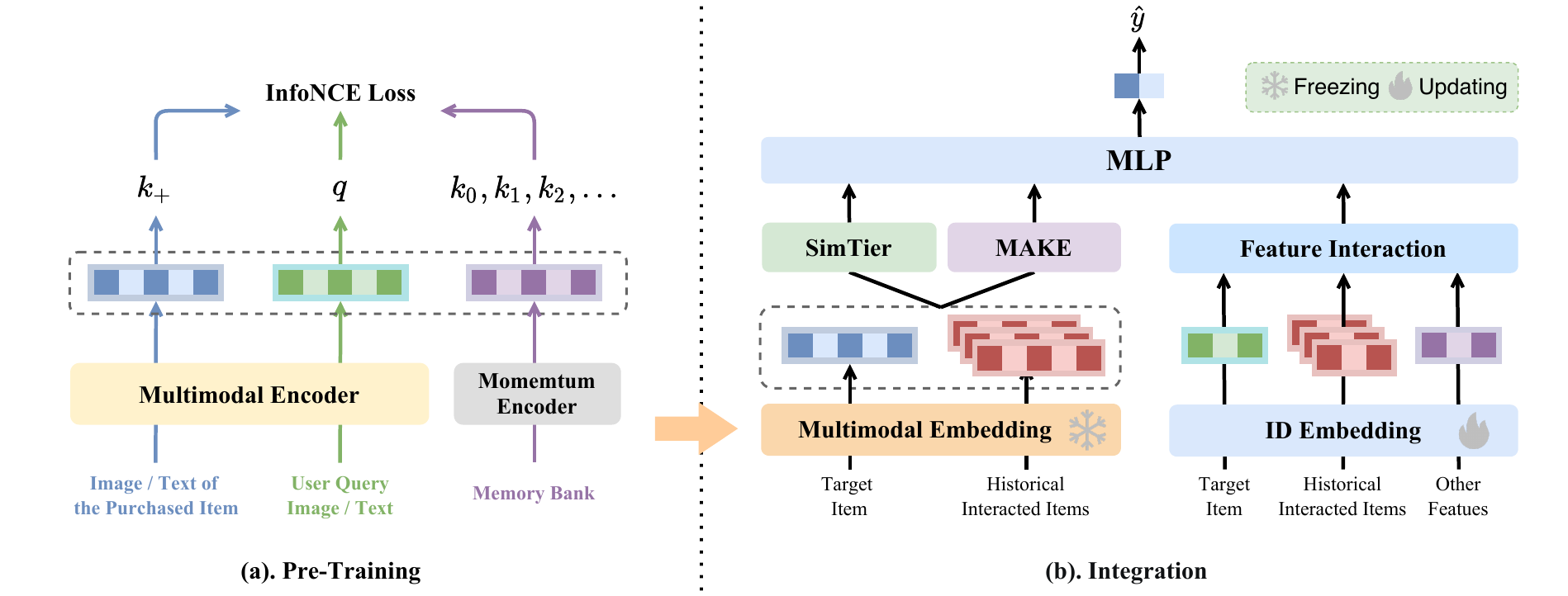} 
\caption{An overview of our two-phase framework: the pre-training of multimodal representations, followed by the integration of pre-trained representations into recommendation models. In the first phase (refer to Figure (a)), we undertake pre-training through semantic-aware contrastive learning. This method equips the multimodal representations with ability to identify semantic similar items.  Subsequently, in the second phase (refer to Figure (b)), we introduce our proposed \textsc{SimTier} and \textsc{make} methods to effectively incorporate the pre-trained multimodal representations into the recommendation models.} 
\label{fig:framework} 
\end{figure*} 

Despite these advancements, a significant number of industrial systems still depend exclusively on ID features. This is often attributed to the concern that the performance gains from multimodal data might not compensate for the costs involved in their deployment. These costs encompass pre-training multimodal encoders, incrementally generating representations for new items, and other necessary upgrades to both online servers and near-line training systems.
Therefore, the successful integration of multimodal representations hinges on the ability to boost their performance benefits while concurrently minimizing deployment costs.  

To accomplish these two key objectives, three practical challenges should be addressed:
\begin{itemize}[leftmargin=*]
    \item \textbf{Design of the pre-training task.} The effectiveness of multimodal representations in enhancing performance hinges on their ability to provide meaningful semantic information, which is difficult for ID features to capture. It is essential to design pre-training tasks that enable multimodal representations to encapsulate such semantic information.
    \item \textbf{Integration of multimodal representation.} 
    The inherent discrepancies between ID features and multimodal representations, such as difference in training epochs, calls for approaches that can effectively incorporating multimodal representations into the ID-based model. These approaches should leverage the strengths of each feature type to enhance the model's overall performance.
    \item \textbf{Design of the production system.} The entire production workflow, including the generation of multimodal representations for new items and their up-to-date application in downstream tasks, should be designed with efficiency.
\end{itemize}

To address these challenges, we adopt a two-phase framework as depicted in Figure~\ref{fig:framework}. Specifically, during the pre-training phase, we propose the \textbf{S}emantic-aware \textbf{C}ontrastive \textbf{L}earning (SCL) method. In this phase, we utilize the user's search query and subsequent purchase action to construct semantically similar sample pairs, capturing the dimensions of semantic similarity that are most relevant to users in e-commerce scenarios. For negative samples, we draw from a large memory bank. The SCL method enables multimodal representations to effectively measure semantic similarities among items.

Upon obtaining high-quality multimodal representations, we propose two approaches to incorporate these representations into the existing ID-based model. Firstly, we develop an approach named {\simtier} to measure the degree of similarity between the target item and items the user has previously interacted with. The resulting {\simtier} vector is then concatenated with other embeddings and fed into the subsequent layers. Furthermore, to address the discrepancy in training epochs between multimodal representations and ID embeddings, we introduce the \textbf{M}ultimod\textbf{A}l \textbf{K}nowledge \textbf{E}xtractor ({\make}) module. The {\make} module separates the optimization of parameters associated with multimodal representations from those of the ID-based model, enabling more effective learning for the parameters related to multimodal representations.

We also present the design of our production system that facilitates the deployment of multimodal representations. Specifically, the system generates multimodal representations for newly introduced items in real-time and ensures these representations are immediately available for the training infrastructure and the online prediction server. This design achieves minimal latency—merely a few seconds between the introduction of an item and the model's use of its multimodal representation. Since mid-2023, multimodal representations have been deployed in Taobao display advertising system, leading to significant performance improvements.

\input{sections/2-method-pretrain} 
\input{sections/3-method-integration}

\input{sections/4-deployment}

\section{Experiment}\label{sec:experiment} 
In this section, we delve into a case study that examines the integration of image representations into the CTR prediction model, aiming to provide a comprehensive analysis.
It is notable that our approach is general, capable of accommodating several modal types (such as text and video) and applicable across different stages within the recommendation systems. 
\subsection{Experiment Setup} 
\subsubsection{Datasets.} 
\begin{figure}[!t]
\centering 
\includegraphics[width=.95\columnwidth]{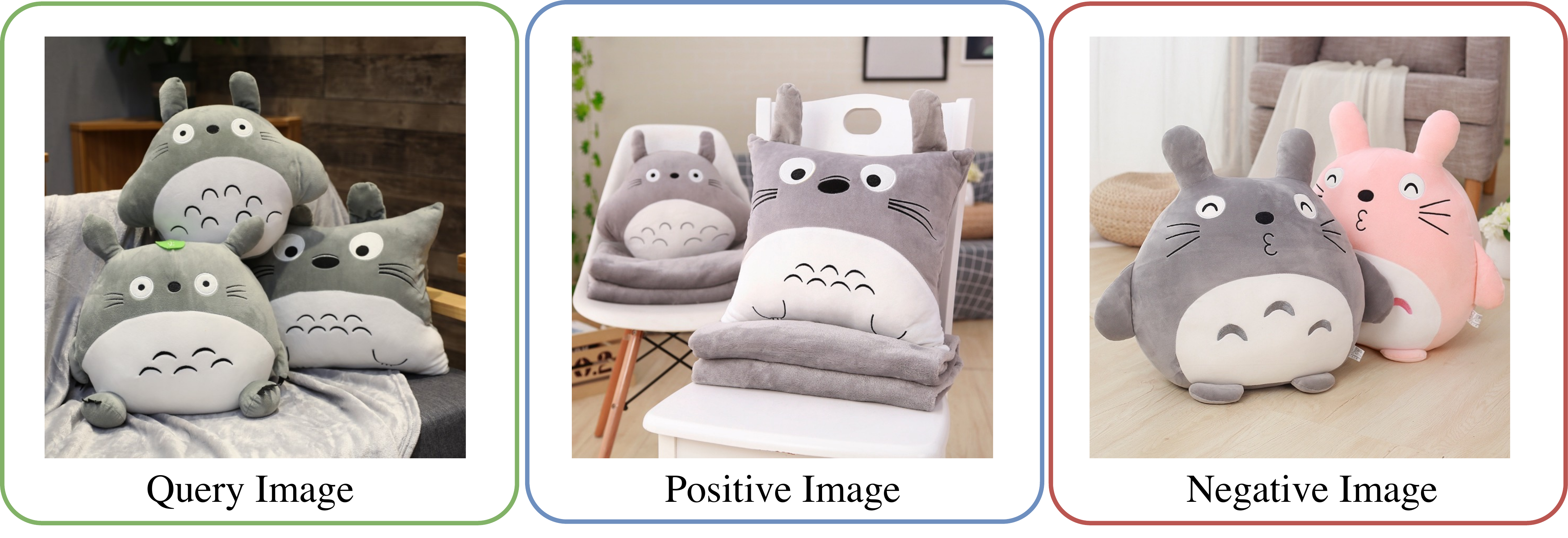}
\caption{A case of the pre-training dataset.}
\label{fig:triplet}
\end{figure}

We start by detailing the datasets used for the pre-training phase and the downstream integration phase.
\begin{itemize}[leftmargin=*] 
    \item \textbf{Pre-training Dataset.} 
    In the pre-training dataset,
    each sample consists of a Query (user's image query) and a Positive (the image of the purchased item). To further enhance the performance, we also add a hard Negative (the image of the clicked item triggered by the positive item). 
    A case of the pre-training dataset is depicted in Figure~\ref{fig:triplet}. 
    \item \textbf{CTR Prediction Dataset.} 
    The CTR prediction dataset is obtained from the Taobao display advertising system, using impression logs of one week. 
\end{itemize}


\subsubsection{Compared Pre-training Methods}\label{sec:compared_pretraining_method}We employ a range of widely-used pre-training method for comparison.
\begin{itemize}[leftmargin=*] 
    \item \textbf{CLIP-O.} CLIP-O refers to the CLIP visual encoder (CLIP-ViT-B/16) that has been pre-trained using a universal dataset~\cite{RadfordKHRGASAM2021Clip}.
    \item \textbf{CLIP-E.} CLIP-E is the fine-tuned version based on the CLIP-O model in the e-commerce scenario using aligned item descriptions and item images.
    \item \textbf{SCL.} The proposed semantic-aware pre-training method. 
    The SCL approach employs Momentum Contrast (MoCo) to expand the set of negative samples. Furthermore, SCL applies triplet loss~\cite{SchroffKP2015FaceNet} to each <Query, Positive, Negative> triplet to effectively discriminate hard negatives. Our experimental analysis investigate the impact of excluding triplet loss and MoCo to assess their contributions to the overall performance.
\end{itemize}
\subsubsection{Compared Recommendation Methods}
We employ a range of widely-used integration approach for comparison.

\begin{itemize}[leftmargin=*] 
    \item ID-based Model (production baseline). The baseline is a ID-based model that underpins our online system. 
    \item Vector. The Vector method utilize the pre-trained multimodal representations as the side-information of each item, and concatenated them with other ID embeddings within the model.
    \item SimScore. Similarity Score (SimScore) can be seen as a simplified version of the Vector method. The semantic similarity score for each historical interacted item with respect to the target item is used as side information.
    \item {\simtier} and {\make}. The proposed {\simtier} and {\make} approaches. 
\end{itemize}

\subsubsection{Evaluation Metrics}
We evaluate both the pre-training performance and CTR prediction performance of the proposed methods. 
\begin{itemize}[leftmargin=*] 
    \item \textbf{Evaluating Pre-training Methods.} 
    Throughout our extensive experimentation, we discovered that the 
    Top-N accuracy (Acc@N) is well correlated with the performance of downstream recommendation models.
    In detail, the Acc@N metric quantifies the ability of the representation to identify semantic similar items:
    \begin{equation}
        \text{Acc@N} = \frac{1}{D} \sum_{i=1}^{D} \mathbb{I}(p_i \in \text{Top}_{N}(q_i, S)),
    \end{equation}
    where $D$ denotes the size of test set. The terms $q_i$ and $p_i$ correspond to the query and the positive of the $i$-th sample, respectively. $S = \{pos_i\}_{i=1}^{D}$ represents the set comprising all positives, and $\text{Top}_N$ is a function that retrieves the top-N results for each query by leveraging a multimodal representation from the set $S$. The symbol $\mathbb{I}(\cdot)$ signifies an indicator function that yields a value of 1 when the $i$-th $p_i$ is among the retrieval results for $q_i$, and 0 in all other cases. 

    \item \textbf{Evaluating Recommendation Methods.}
    We assess the effectiveness of the CTR prediction model using the AUC 
    and Group AUC (GAUC) metrics, where a higher AUC/GAUC value signifies superior ranking ability~\cite{ZhuJTPZLG2017OCPC,zhou2018din,ShengZZDDLYLZDZ2021STAR}. 
\end{itemize}
\subsection{Performance on Pre-Training Dataset}
\subsubsection{Rationale for Utilizing Accuracy to Evaluate the Effectiveness of Pre-trained Representations}\label{sec:correlation}

The most precise way to gauge the effectiveness of pre-trained representations is by measuring the improvement in recommendation accuracy with the integration of multimodal representations. However, this evaluation process can be lengthy for iteration of pre-training methods, and an intermediary metric for a quicker assessment of pre-trained multimodal representations is desirable.

In our research, we observed \textbf{a strong correlation between the enhancement in pre-training accuracy and the boost in recommendation performance}. We illustrate this relationship in Figure~\ref{fig:correlation}, where we can see that the improvement of Acc@1 is consistent with the improvement of GAUC. Hence, we predominantly rely on pre-training accuracy to determine the quality of multimodal representations. The exploration of other potential intermediary metrics for evaluating pre-trained multimodal representations remains a interesting topic for future work.

\subsubsection{Importance of Semantic-Aware Contrastive Learning} \label{sec:exp_semantic_similarity}
To investigate how different pre-training tasks affect the quality of multimodal representations, we conducted a series of experiments.
The results, presented in Table~\ref{tab:pretrain}, offer two important observations. First, the proposed SCL pre-training method surpasses other semantic-similarity-agnostic methods, emphasizing \textbf{the necessity of the semantic-aware learning.}. Second, incorporating techniques like Momentum Contrast (MoCo)~\cite{He0WXG2020Moco} and Triplet Loss~\cite{SchroffKP2015FaceNet} further enhances the quality of the multimodal representations, demonstrating the choice of negative sample greatly impacts the performance.

\begin{table}[!t]
\centering
\caption{Pre-training performance of different methods}
\begin{tabular}{l|cc}
\toprule
Method & Acc@1 & Acc@5 \\
\midrule
CLIP-O & 0.2559 & 0.3575 \\
CLIP-E & 0.2952 & 0.3917 \\
\midrule
SCL & \textbf{0.7474} & \textbf{0.8850}  \\
\quad w/o Triplet Loss & 0.6957 & 0.8604  \\
\quad w/o Triplet Loss \& MoCo & 0.5760 & 0.7590  \\
\bottomrule
\end{tabular}
\label{tab:pretrain}
\end{table}

\subsection{Performance on CTR prediction Dataset}\label{sec:exp_integration}

\begin{table}[!t]
\centering  
\caption{Overall performance on CTR prediction dataset.}  
\begin{tabular}{l|cc}
\toprule  
Method & GAUC & AUC  \\
\midrule  
ID-based Model & - & -   \\
Vector & +0.29\% & +0.18\%  \\
SimScore & +0.77\% & +0.40\%  \\
\midrule  
{\simtier} & +0.96\% & +0.59\%  \\
{\make} & +0.93\% & +0.51\% \\
\midrule 
{\simtier}+{\make} & \textbf{+1.25\%} & \textbf{+0.75\%} \\
\bottomrule  
\end{tabular}
\label{tab:overall}
\end{table}

\begin{figure*}[!tbp]
\centering
\subfigure[]{
\includegraphics[width=.315\linewidth]{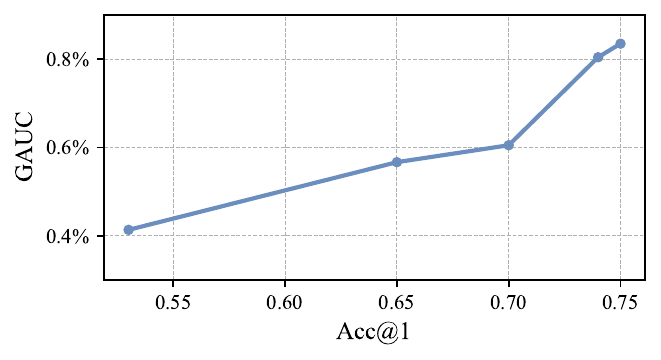}
\label{fig:correlation}
}
\subfigure[]{
\includegraphics[width=.32\linewidth]{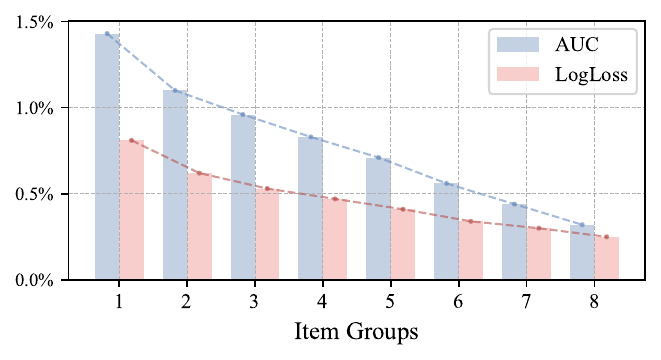}
\label{fig:long-tail}
}
\subfigure[]{
\includegraphics[width=0.315\linewidth]{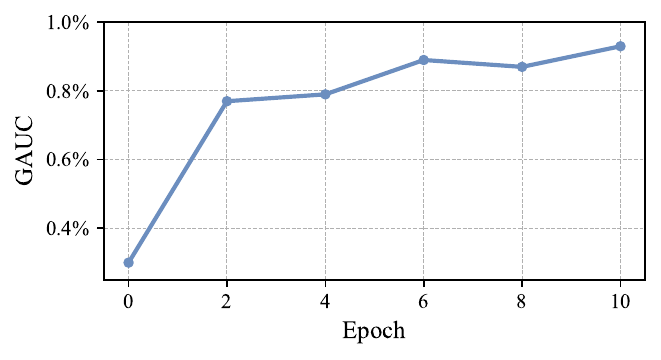}\label{fig:ablation-make}
}
\caption{(a) The correlation between pre-training metric and CTR prediction performance. (b) The relative improvement on item groups with different frequency.
(c) The relative performance of the recommendation model with different pre-training epochs on {\make}. Note that 0 epoch implies that {\make} does not undergo pre-training, but is instead jointly optimized with the downstream task. 
}
\end{figure*}

\subsubsection{
Performance on Different Integration Strategies}\label{sec:approaches}
In the CTR prediction dataset, we evaluate the proposed {\simtier} and {\make} against other methods. The overall results are shown in Table~\ref{tab:overall}, from which two observations can be noted. Firstly,  {\simtier} and {\make} outperform other methods significantly. 
Secondly, the combination of {\simtier} and {\make} can further improve the performance, with a 1.25\% increase in GAUC, 0.75\% increase in AUC compared with the ID-based model. The above results demonstrate the effectiveness of the proposed methods on integrating multimodal representations into the ID-based model.




\subsubsection{Performance on Different Training Epochs of {\make}}


To explore the impact of multi-epoch pre-training of {\make} on the final recommendation performance, we conduct experiments where {\make} is pre-trained for varying numbers of epochs before integration into the recommendation model. The results are shown in Figure~\ref{fig:ablation-make}. Note that 0 epochs implies that {\make} does not undergo pre-training and is instead jointly optimized with the downstream task. The results indicate that as the number of pre-training epochs for {\make} increases, the performance of the final recommendation model also improves. This demonstrates that multi-epoch training of {\make} effectively enhances model performance.

\subsubsection{Performance on Infrequent Items.} 
To examine the generalization ability of multimodal representation on long-tail items, we assess the relative improvement across item groups categorized by their frequency of occurrence. We divide all items into eight groups, with Group 1 containing items of the lowest frequency and Group 8 encompassing those with the highest frequency in the training dataset. Afterward, we compute the relative improvement of AUC as |AUC$_{\text{MM}}$-AUC$_{\text{ID}}$|/AUC$_{\text{ID}}$ for each group, where $MM$ denotes the combination of  {\simtier} and {\make} method and $ID$ represents the ID-based production baseline model. We also compute the relative improvement of LogLoss as |LogLoss$_{\text{MM}}$-LogLoss$_{\text{ID}}$|/LogLoss$_{\text{ID}}$ to measure the calibration ability~\cite{GuoPSW2017OnCalibration,ShengGCYHDJXZ2023JRC}. The result presented in the Figure~\ref{fig:long-tail} indicates that the multimodal representation exhibits a significant improvement in all groups, demonstrating the effectiveness of our model over different types of items. Meanwhile, we see \textbf{a more significant improvement for low-frequency items}. The result demonstrate that multimodal representations can address the shortcomings of ID-based model on long-tail items and enhance the prediction accuracy.



\subsection{Online Performance}\label{sec:online}


Since mid-2023, multimodal representations have been integrated into pre-ranking, ranking, and re-ranking models within the Taobao display advertising system, resulting in substantial performance improvements. For instance, incorporating image representations in the CTR prediction model yielded a overall 3.5\% increase in CTR, a 1.5\% boost in RPM, and a 2.9\% rise in ROI. Notably, the impact was even more pronounced for new ads (created within the last 24 hours), with improvements of 6.9\% in CTR, 3.7\% in RPM, and 7.7\% in ROI. The significant gains on new ads also validate the effectiveness of multimodal data in mitigating the cold-start issue.

\input{sections/6-related} 
\section{Conclusion and Discussion}\label{sec:conclusion} 
Multimodal-based recommendation has attracted attention over decades. 
However, integrating multimodal representations into industrial systems presents many hard challenges, particularly in the realms of representation quality, integration methods, and system implementation—challenges that are amplified within the context of large-scale industrial systems. 

In this study, we delve into these challenges and share the approaches we employed for pre-training and incorporation of multimodal representations. Additionally, we provide insights gleaned from our experiences during the online deployment stage. We believe the strategies and insights we have amassed through our journey will serve as a valuable resource for those aiming to expedite the adoption of multimodal-based recommendations in industrial systems.
\bibliographystyle{ACM-Reference-Format}
\bibliography{sample-base}



\end{document}

%% file: sections/2-method-pretrain.tex
\section{Preliminaries}\label{sec:preliminary}
Before delving into the specifics, we first introduce the typical ID-based model structure utilized in different stages (including retrieval~\cite{zhu2018tdm,HuangSSXZPPOY2020FacebookEBR}, pre-ranking~\cite{WangZJZZGCold,ZhaoGZHLSWZJXZ2023PrerankLTR}, and ranking stages~\cite{CovingtonAS2016YouTubeDNN,zhou2018din,BianWRPZXSZCMLX2022CAN,GuSFZZ2021Defer,chan2023capturing}) of industrial system. 

\textbf{ID Features in Recommendation Models:} Common recommendation models are trained on large-scale datasets comprising billions of ID features~\cite{Jiang2019XDL,zhang2022picasso}. These ID features serve to represent users profiles, user historical interacted items, the target item (to be predicted), and the contextual information. For example, we can represent the target item with its respective item ID and category ID and represent user historical interacted items through a sequence of corresponding item IDs and category IDs. 

\textbf{Structure of ID-based Model:} The ID-based recommendation model follows an embedding and MLP (Multi-Layer Perceptron) architecture, which typically incorporates \textit{historical behavior modeling} modules~\cite{zhou2018din,zhou2019dien}.  Initially, all ID features are converted into embeddings.  The historical behavior modeling modules then measures a user's interest towards the target item by analyzing the relevance between the embedding of target item and embeddings of the user's historical interacted items. Specifically, the modules produce fixed-length vectors by aggregating the embeddings of the target item and those of the historical interacted items. These vectors are then concatenated with other ID embeddings to form the input for subsequent MLP, which produce the final prediction.

\section{Pre-Training of Multimodal Representations}\label{sec:pretraining} 

As for the multimodal data, its utilization can improve historical behavior modeling. Specifically, multimodal data can be used to measure the semantic similarity between the target item and users' historical interacted items. Take item images as an example, they can be used to measure the visual similarity between the image of the target item and those of historical interacted items. Intuitively, a higher semantic similarity indicates a stronger resemblance between the target item and the users' historical behavior, suggesting a higher likelihood of the user's interest. This insight emphasizes the importance of designing a pre-training task tailored to enable multimodal representations to effectively discern the semantic similarity across item pairs.


\subsection{Semantic-Aware Contrastive Learning}\label{sec:semantic-pretraining}

To derive representations with the capability to discern semantic similarity, we propose the semantic-aware contrastive learning (SCL) method that attracts the semantically similar sample pairs and repulses the dissimilar sample pairs. To accomplish this, it is essential to define semantically similar and dissimilar pairs for supervision.
To understand the importance of this point, consider the example shown in Figure~\ref{fig:triplet}, where the three pillows are almost identical yet display slight variances, like differences in patterns or minor appearance discrepancies. If the definition of semantically similar pairs is inadequate, the representation might fail to capture such subtle differences. Indeed, these slight distinctions are frequently missed by representations focused on capturing general concepts.

In the following, we will elaborate our construction of pre-training dataset tailored for e-commerce scenario and the optimization strategies of contrastive learning process.

\subsection{Construction of Pre-Training Dataset}\label{sec:pre-train-dataset}

In the context of e-commerce, \textbf{a user's search query and subsequent purchase action} often signifies a strong semantic similarity between the query and the purchased item. For example, if a user searches for an image of a pillow and subsequently purchases a pillow, this sequence of actions indicates that the two images (the queried image and the image of the purchased item) are semantically similar enough to satisfy the user's purchase intentions. Thus, as shown in Table~\ref{table:pre_training_pair}, in training the text encoder, we pair the text of the user's search query with the title of the item they ultimately purchased as the semantically similar pair. Similarly, for the image modality, user's  image query (obtained from the image search scenario of Taobao) is paired with the image of the subsequent purchased item. This pairing strategy naturally captures the dimensions of semantic similarity that are most relevant to users in e-commerce scenarios, reflecting the elements that influence their purchasing decisions.

For each semantically similar pair, we consider all other samples as potential dissimilar samples, which can be achieved by using the samples in the current mini-batch as negatives. To further improve the model performance, we aim to increase the number of negative samples available during training. Specifically, we draw inspiration from MoCo~\cite{He0WXG2020Moco} and adopt a technique that updates the model with momentum, facilitating sampling more negatives from a larger memory bank. 
More sophisticated strategies for constructing negative pairs, such as identifying hard negatives~\cite{SchroffKP2015FaceNet}, will be elaborated in Section~\ref{sec:compared_pretraining_method}.



\begin{table}[!t]
\centering
\caption{The construction of semantically similar pair for pre-training.}
{
\begin{tabular}{c|c}
    \toprule 
    modality   & semantically similar pair  \\
    \midrule
    Image
    & <user's image query, image of the purchased item>
    \\
    \midrule
    Text & <user's text query, title of the purchased item>  \\
    \bottomrule 
\end{tabular}
}
\label{table:pre_training_pair}
\end{table}

\begin{figure*}[!t]
\centering 
\includegraphics[width=.9\textwidth]{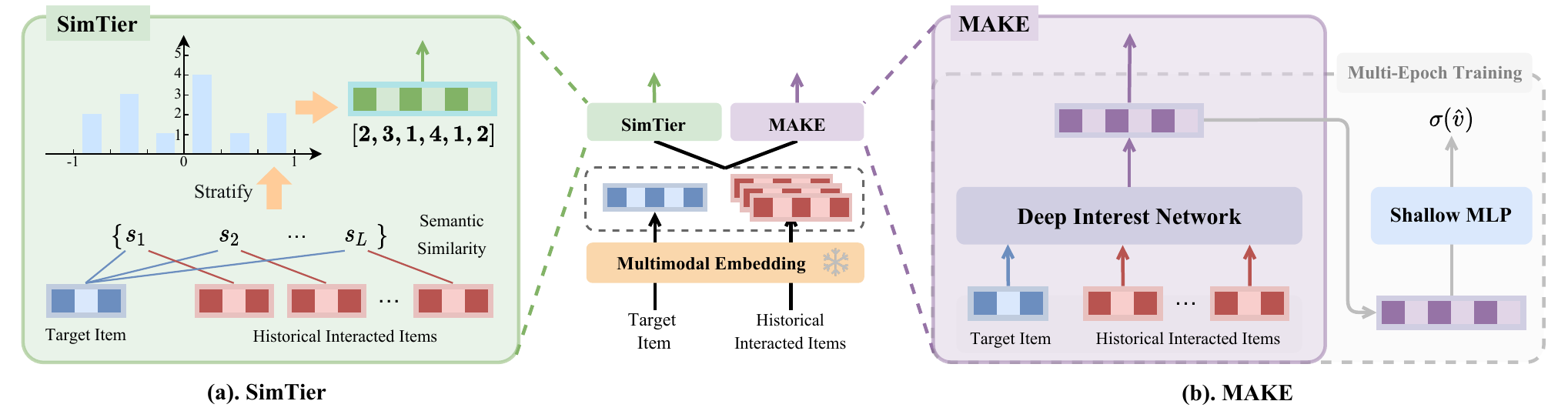} 
\caption{An illustration of the proposed \textsc{SimTier} and Multimodal Knowledge Extractor (\textsc{MAKE}) approaches, with details provided in Section~\ref{sec:simtier} and \ref{sec:make}, respectively.} 
\label{fig:simtier}
\end{figure*}

\subsection{Optimization}\label{sec:optimization}



We utilize the well-regarded InfoNCE loss~\cite{oord2018representation} as the loss function. Given an encoded query $q$ and its corresponding encoded positive sample $k$, along with ${k_0, k_1, \dots, k_K}$ representing the set of encoded sample representations in the memory bank, where $K$ denotes the memory bank size~\cite{He0WXG2020Moco}, the InfoNCE loss employs the dot product to measure similarity (with all representations being L2 normalized). As shown in Equation~\ref{eq:infonce}, the loss value decreases when query $q$ closely matches its designated positive sample $k$ and diverges from all other samples within the memory bank. 
\begin{equation}
L_{\text{InfoNCE}} = -\log\frac{\exp(q\cdot k_{+})/\tau }{\sum_{i=0}^{K}\exp(q\cdot k_{i})/\tau }.
\label{eq:infonce}
\end{equation}
Here, $\tau$ represents a learnable temperature parameter. For our experiments, we set the value of $K$ to 196,800.

By this means, the SCL method enables the representations to have the ability to discern fine differences between comparable items, which are crucial for recommendation models.



%% file: sections/3-method-integration.tex
\section{Integration with Recommendation Models}\label{sec:integration}
A direct method to integrate multimodal representations into an ID-based recommendation model involves concatenating these multimodal representations with the ID embeddings for both the target item and users' past interacted items~\cite{GeZZCLYHLSLYHZZ2018CrossMedia,Yang2023Courier}. This concatenation is followed by the utilization of user behavior modeling modules, which are subsequently input into the MLP for the final prediction.
Although this method is straightforward, we find that the performance improvements are modest. To investigate this matter, we share our observations and insights. 


\subsection{Observations and Insights} 
\label{sec:insight}
We begin by sharing our observations and insights on incorporating multimodal representations.

\textbf{Observation 1: simplifying the usage of multimodal representations improves performances.} 
Our research reveals that the direct integration of multimodal representations into the ID-based model does not yield optimal performance, as explained in more detail in Section~\ref{sec:exp_integration}. This issue arises because the parameters associated with multimodal representations, e.g., the parameters of the MLP that are connected with multimodal representations, are not adequately learned during the joint training process with the ID embeddings~\cite{Yang2023Courier}. In contrast, strategies that simplify the usage of multimodal representations, for instance, transforming them into semantic IDs (thereby representing the embedding vectors with IDs)~\cite{Singh2023SemanticID,Yang2023Courier}, appear to offer improved performance. 

\textbf{Observation 2: ID-based and multimodal-based models have training epoch discrepancy.} 
In industrial scenario, ID-based models are typically trained for only one epoch to avoid overfitting~\cite{ZhangSZJHDZ2022OneEpoch}. In contrast, models in CV and NLP field often undergo training over multiple epochs.
A natural question arises: how many training epochs are ideal for a multimodal-based recommendation model? To answer this question, we developed a recommendation model that exclusively utilizes multimodal representations as input features, without any ID features, and analyzed how its performance varied with the number of training epochs.

\begin{figure}[!t]
\centering 
\includegraphics[width=0.9\columnwidth]{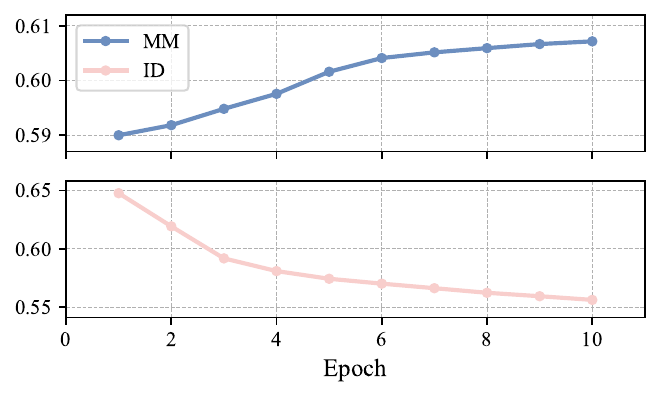}
\caption{The multimodal-based CTR prediction model (MM) demonstrates a continuous increase in GAUC after several training epochs. In contrast, the ID-based model (ID) shows a sharp decline in GAUC during testing after the second epoch of training.}
\label{fig:multi-epoch}
\end{figure}

The detailed convergence curve is illustrated in Figure~\ref{fig:multi-epoch}. We find that model leveraging multimodal data benefits from training across multiple epochs on the same dataset, with its performance showing notable enhancements as the number of epochs increases. In contrast, ID-based models suffer from the one-epoch overfitting phenomenon, where model performance dramatically degrades at the beginning of the second epoch~\cite{ZhangSZJHDZ2022OneEpoch}. 
The result suggests that the parameters associated with multimodal representations require more epochs to converge properly, which contrasts with the behavior of the ID-based model.
\textbf{Consequently, when incorporating multimodal representations into an ID-based model, and trained over only one epoch, there is a risk that the multimodal-related parameters may not be sufficiently trained.}


\subsection{Method \uppercase\expandafter{\romannumeral 1}: {\simtier}}
\label{sec:simtier} 

The observation 1 calls for simplifying the usage of multimodal representations. To this end, we propose a straightforward yet effective method  {\simtier}. As shown in Figure~\ref{fig:simtier} (a), {\simtier} begins by computing the dot product similarity between the multimodal representation of the target candidate item, denoted as $v_c$, and the multimodal representations of the user's historically interacted items $\{v_i\}_{i=1}^{L}$,
\begin{equation} 
\begin{aligned} 
s_i = v_i \cdot v_c, \forall i\in\{1,\dots,L\}.
\end{aligned} \label{eq:cosine} 
\end{equation} 



Following the calculation of the similarity scores, we partition the score range of [-1.0, 1.0] into $\bf{N}$ predefined tiers. Within each tier, we count the number of similarity scores that fall into that corresponding range. Hence, we obtain an $\bf{N}$-dimensional vector, with each dimension representing the number of similarity scores in the corresponding tier. Thus, {\simtier} effectively converts a set of high-dimensional multimodal representations into a $\bf{N}$-dimensional vector that encapsulate the degree of similarity between the target item and the user's historical interactions.
The obtained $\bf{N}$-dimensional vector is then concatenated with other embeddings and fed to the following MLP. 
We provide the pseudo code of  {\simtier} in Algorithm~\ref{alg:simtier}.

\subsection{Method \uppercase\expandafter{\romannumeral 2}: Multimodal Knowledage Extractor (\textsc{MAKE})}
\label{sec:make} 

\begin{algorithm}[!t]
\caption{A Tensorflow-style Pseudocode of the SimTier.}
\label{alg:simtier}
\definecolor{codeblue}{rgb}{0.25,0.5,0.5}
\definecolor{codekw}{rgb}{0.85, 0.18, 0.50}
\begin{lstlisting}[language=python]
# B: batch size, S: sequence length, N: tier number, D: dim
# Input: target [B, D], seq [B, S, D]
# Output: sim_tier [B, N]

# Compute the similarity scores [B, S] 
sim_score = reduce_sum(expand_dims(target, 1)*seq, axis=2)
# Assign tier to each score [B, S]
indices = reshape(ceil((sim_score + 1) / 2 * N), [-1, S])
# Accumulate counts for each tier [B, N]
weight = equal(
            reshape(range(0, N, 1), [1, N, 1]),
            expand_dims(indices, axis=1)
        )
sim_tier = reduce_sum(weight, axis=2, keep_dims=True)
\end{lstlisting}
\end{algorithm}



To address the difference in training epochs required for ID features versus multimodal representations, we introduce the \textbf{M}ultimod\textbf{A}l \textbf{K}nowledge \textbf{E}xtractor ({\make}) module, decoupling the optimization of multimodal related parameters from that of ID features.  The {\make} module consist of two steps: 1) multi-epoch training to extract useful multimodal knowledge and 2) knowledge utilization by the downstream task.

\textbf{Multi-epoch training of multimodal related parameters.} 
The goal of the {\make} module is pre-training the parameters related to multimodal representations over multiple epochs to ensure their convergence. In practice, we utilize the CTR prediction task as the recommendation pre-training task. 
As shown in Figure~\ref{fig:simtier} (b), we first develop a  DIN-based user behavior modeling module~\cite{zhou2018din}. This module processes the pre-trained multimodal representations of  the target item and historical interacted items, resulting in the output $\bf{v}_\text{MAKE}$:
\begin{equation} 
\begin{aligned} 
\bf{v}_\text{MAKE} = \text{DIN}(\{\bf{v}_i\}_{i=0}^{L}, \bf{v}_c). 
\end{aligned} \label{eq:mic_encoder} 
\end{equation}


After that, $\bf{v}_\text{MAKE}$ is fed into a four-layer Multi-layer Perceptron (MLP$_\text{MAKE}$) and produced the logit $\hat{v}$ 
\begin{equation} 
\begin{aligned} 
\hat{v} = \text{MLP}_\text{MAKE}(\bf{v}_\text{MAKE}). 
\end{aligned} \label{eq:mic_train} 
\end{equation} 
Then we optimize the cross-entropy loss between the predicted click probability and the binary click label $y$, as shown in Equation~\ref{eq:mic_loss}.
\begin{equation} 
\begin{aligned} 
\mathcal{L}_{\text{MAKE}} = &\sum  -y \log \sigma(\hat{v}) - (1-y) \log (1-\sigma(\hat{v}))
\end{aligned} \label{eq:mic_loss} 
\end{equation} 

The recommendation pre-training task allows the {\make} module to refine its parameters via training over multiple epochs and extract knowledge $\bf{v}_\text{MAKE}$ from multimodal representations, thereby enhancing its effectiveness for recommendation tasks. 

\textbf{Knowledge utilization.} 
After acquiring the vector $\bf{v}_\text{MAKE}$, the subsequent step is to integrate it into the downstream recommendation task. Practically, we concatenate $\bf{v}_\text{MAKE}$ and intermediate outputs from MLP$_\text{MAKE}$ with other embeddings and input this combined data into the subsequent layers. 
The implementation of the multi-epoch training of {\make} module reconciles the training epochs difference required for ID embeddings and multimodal representations, resulting in better performances.


%% file: sections/4-deployment.tex
\section{INDUSTRIAL DESIGN for Online Deployment}\label{sec:deployment}

In industrial settings, new items (including advertisements) are constantly being created. To maintain prediction accuracy for these new items, it's crucial for the recommendation model to acquire the multimodal representation of new items in real-time. This calls for the ability of continuous generating multimodal representations for new items and a real-time utilization by near-line trainer and online server. 
\begin{figure}[!t]
\centering 
\includegraphics[width=\columnwidth]{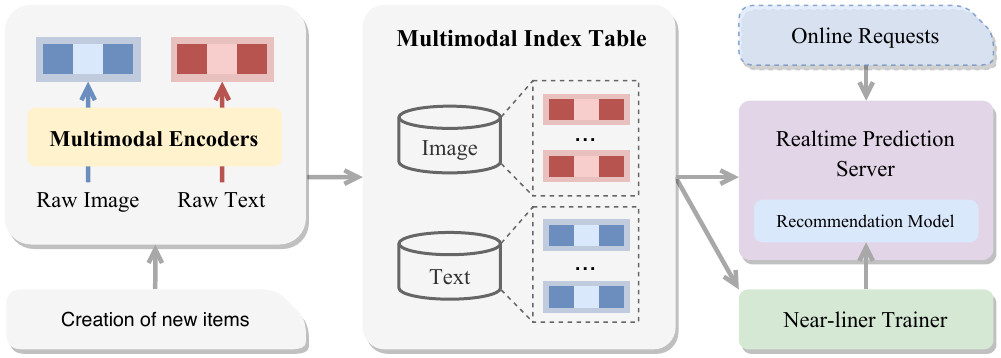} 
\caption{An overview of the online system.} 
\label{fig:production} 
\end{figure} 
The illustrative overview of our online system is provided in Figure~\ref{fig:production}. To achieve the real-time generation of  multimodal representations,  upon the introduction of new items, the system automatically initiates a request to the pre-trained multimodal encoders to compute the multimodal representations for these new items.  Once inferred, these representations are sent to the multimodal index table.  Following this step, the downstream training systems and inference servers are able to retrieve the multimodal representations from the index table, facilitating near-line training and real-time online prediction capabilities. This process ensures minimal latency—reduced to just a few seconds—between an item's introduction and the utilization of its multimodal representation by the model.

%% file: sections/6-related.tex
\section{Related Work}\label{sec:related}  
Currently, ID features constitute the core of industrial recommendation models~\cite{cheng2016wide,CovingtonAS2016YouTubeDNN,zhou2018din,zhou2019dien,ZhangSZJHDZ2022OneEpoch,chan2020selection,zhang2022keep,hu2023ps}. Despite their widespread adoption, ID features have notable limitations, including the challenge of capturing semantic information and the persistent issue of the cold-start problem~\cite{ScheinPUP2002ColdStart,GeZZCLYHLSLYHZZ2018CrossMedia,MoLXLJ2015ImageColdStart,wu2022adversarial}. In contrast, multimodal data offer rich semantic information, prompting numerous studies to explore their incorporation into recommendation models. Some research has investigated the potential of learning multimodal representations in an end-to-end manner alongside recommendation model training~\cite{ChenSLLH2016DeepCTRMM,YuanYSLFYPN2023WhereToGo,Elsayed2022EndtoendImageFashion}. However, the substantial computational resources required for such processes often preclude their adoption in industrial systems. Therefore, our focus is on the two-phase paradigm~\cite{ChengZAMZZN2012MMfea4CTR,LynchAA2016ImageDontLie,GeZZCLYHLSLYHZZ2018CrossMedia,HeM2016UpsDownsVisual,Singh2023SemanticID,Yang2023Courier,PalEZZRL2020PinnerSage}, and we aim to share our methods and the valuable insights. 
